\documentclass[twocolumn,superscriptaddress,
 amsmath,amssymb,
 aps,
 prl,
]{revtex4-2}
\usepackage{amsmath,amssymb}
\usepackage{graphicx,epsfig}
\usepackage[greek,english]{babel}
\usepackage{bbold}
\usepackage{comment}
\usepackage{soul}             

\usepackage{mathtools}
\usepackage{makecell,tabularx}
\usepackage{xcolor}
\setcellgapes{1pt}
\usepackage{subfigure} 
\makegapedcells

\makeatletter\AtBeginDocument{\let\@elt\relax}\makeatother

\begin{document}
\bibliographystyle {plain}
\pdfoutput=1
\def\oppropto{\mathop{\propto}} 
\def\opsimeq{\mathop{\simeq}}
\def\opoverderline{\mathop{\overline}}
\def\operarrow{\mathop{\longrightarrow}}
\def\opsim{\mathop{\sim}}

\def\opmin{\mathop{\min}} 
\def\opmax{\mathop{\max}} 
\def\oplim{\mathop{\lim}}

\title{A universal property of random trajectories in bounded domains}
\author{T. Binzoni}
\affiliation{Department of Radiology and Medical Informatics, University Hospital, Geneva, 1211, Switzerland}
\email{tiziano.binzoni@unige.ch}
\author{E. Dumonteil}
\affiliation{Institut de Recherche sur les Lois Fondamentales de l’Univers, CEA, Universit\'e Paris-Saclay, 91191 Gif-sur-Yvette, France}
\email{eric.dumonteil@cea.fr}
\author{A. Mazzolo}
\affiliation{ 
Universit\'e Paris-Saclay, CEA, Service d'\'Etudes des R\'eacteurs et de Math\'ematiques Appliqu\'ees, 91191, Gif-sur-Yvette, France}
\email{alain.mazzolo@cea.fr}

\begin{abstract}
The celebrated invariance property states that particles entering a bounded domain, with isotropic and uniform incidence,
spend on average $\langle \ell \rangle=4V/S$ length inside, no matter how they scatter. We show that this remarkable property is merely the infinite-length limit of an even broader law: for any curves randomly placed and oriented in space -stochastic or deterministic, generated by ballistic or diffusive dynamics, with possible stopping or branching, in two or more dimensions-
$ \displaystyle \frac{1}{\langle \ell \rangle}= \frac{1}{\langle L\rangle}+ \frac{1}{\langle \sigma \rangle} $, 
with $\langle\ell\rangle$ its mean in-domain path, $\langle L\rangle$ its mean total length, and $\langle\sigma\rangle$ the mean chord of the domain, a known geometric quantity related to the volume-to-surface ratio.  Derived solely from the kinematic formula of integral geometry, the result is independent of step-length statistics, memory, absorption, and branching, making it equally relevant to photons in turbid tissue, active bacteria in micro-channels, cosmic rays in molecular clouds, or neutron chains in nuclear reactors.  Monte-Carlo simulations spanning straight needles, Y-shapes, and isotropic random walks in 2D and 3D confirm the universality and demonstrate how a local measurement of $\langle \ell \rangle$ yields $\langle L\rangle$ without ever tracking the full trajectory.
\end{abstract}

\maketitle

\emph{Introduction} ---  
The statistics of path‑lengths in bounded domains underpin a broad range of transport phenomena, from neutron diffusion to the foraging of micro‑organisms. A celebrated result, commonly known as the \textit{invariance property} (IP), states that for particles entering a bounded three-dimensional domain with an isotropic and uniform incidence~\cite{footnote_1}, the average path length within the domain, $\langle \ell \rangle$, is given by
\begin{equation}
\label{eq_IP}
\langle \ell \rangle = \frac{4V}{S},
\end{equation}
where $V$ and $S$ denote the volume and surface area of the domain, respectively. Remarkably, this result holds independently of the microscopic details of the random walk—mean free path, scattering law, or step‑length distribution~\cite{Bardsley,Blanco_EPL,Mazzolo_EPL,Benichou,ZDM}.  The IP has been rediscovered repeatedly across disciplines, validated through Monte Carlo simulations and verified experimentally in complex media
~\cite{Pierrat,Tommasi,book_Tiziano,Martelli,Tommasi_2024,Savo}.

Yet, real trajectories are rarely the ideal infinite memory‑less walks, implicit in Eq.\eqref{eq_IP}~\cite{Bechinger,Elgeti_other_jump,Davis}. Paths can start or stop inside the volume, or even branch as in active‑particle kinetics and cascade processes.  Whether any vestige of the IP survives in this far broader class of stochastic curves has remained unclear.

Here we show that it does, and in an unexpectedly simple form. 

For an ensemble of random curves with average total length $\langle L\rangle$ and average in-domain path length $\langle \ell\rangle$, we prove the universal identity
\begin{equation}
\label{eq_cauchy_nD_random} 
  \frac{1}{\langle \ell\rangle}= \frac{1}{\langle L\rangle}+ \frac{1}{\langle \sigma\rangle}  ,
\end{equation}
where $\langle \sigma \rangle$ denotes the mean chord length of the domain. In three dimensions, $\langle \sigma \rangle = \frac{4V}{S}$, while in two dimensions, $\langle \sigma \rangle = \frac{\pi S}{P}$, with $P$ the perimeter of the domain.

This relation, hereafter referred to as the generalized Invariance Property, shows that the local observation of path segments suffices to infer the global length of an arbitrary trajectory ensemble, no matter how the curves meander, terminate, or branch—as long as their spatial distribution remains uniform
and isotropic. Notably, the domain can have arbitrary shape—convex or nonconvex, with holes, or even composed of multiple disconnected components as shown in Fig.~\ref{fig-K0andK1-crop}.
{                         
\begin{figure}[h]
\centering
\includegraphics[width=3.3in,height=2.in]{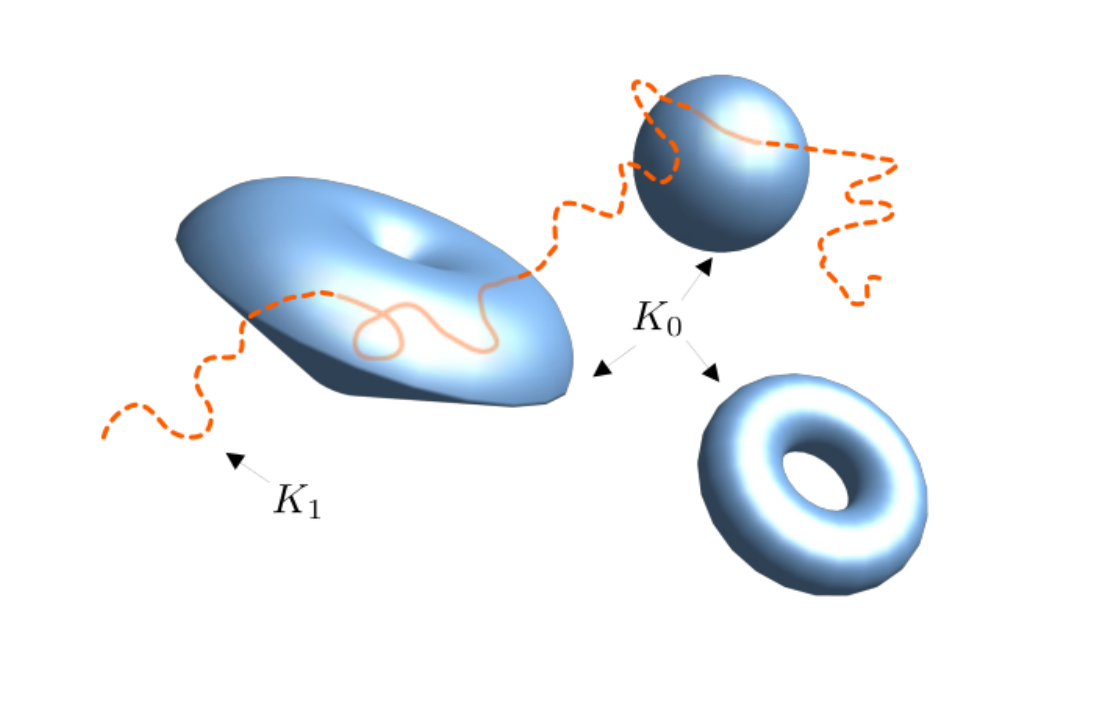}
\setlength{\abovecaptionskip}{-10pt} 
\caption{
An example of a random trajectory traversing a nonconvex domain $K_0$, defined as the union of 3 volumes, one of which contains a hole.
}
\label{fig-K0andK1-crop}
\end{figure}
}

Our derivation dispenses entirely with transport equations.  We treat trajectories as one‑dimensional manifolds and invoke the fundamental kinematic formula of integral geometry, thereby exposing the purely geometric skeleton hidden beneath prior probabilistic proofs.

Below we first present the result for spatial dimension $n \geq 3$, where self‑intersections of the trajectories are atypical.  We then address the more subtle two‑dimensional case, introducing a graph‑theoretic counting scheme that resolves loops and branching vertices.  Finally, Monte‑Carlo simulations covering straight needles, branched Y‑shapes, open triangles, and isotropic random walks confirm Eq.\eqref{eq_cauchy_nD_random} in both 2D and 3D.

\emph{The $n$-dimensional case ($n \geq 3$)} --- 
We consider a collection of independent random particles freely evolving in space ($n\ge 3$) under isotropic and translationally invariant equilibrium conditions. Our goal is to characterize their behavior within a bounded observation domain $K_0 \subset \mathbb{R}^n$, focusing on the average in-domain path length $\langle \ell \rangle$ —or equivalently, the mean residence time if particles move at constant speed.

Two ingredients are required: (i) the total length of all trajectory pieces lying inside the observation window $K_0$, and (ii) the total number of such pieces (i.e., connected components of trajectory intersections with $K_0$). 
Their ratio yields the desired mean path length $\langle \ell \rangle$.

\begin{figure*}[t]
  \centering
  \includegraphics[width=5in,height=3.5in]{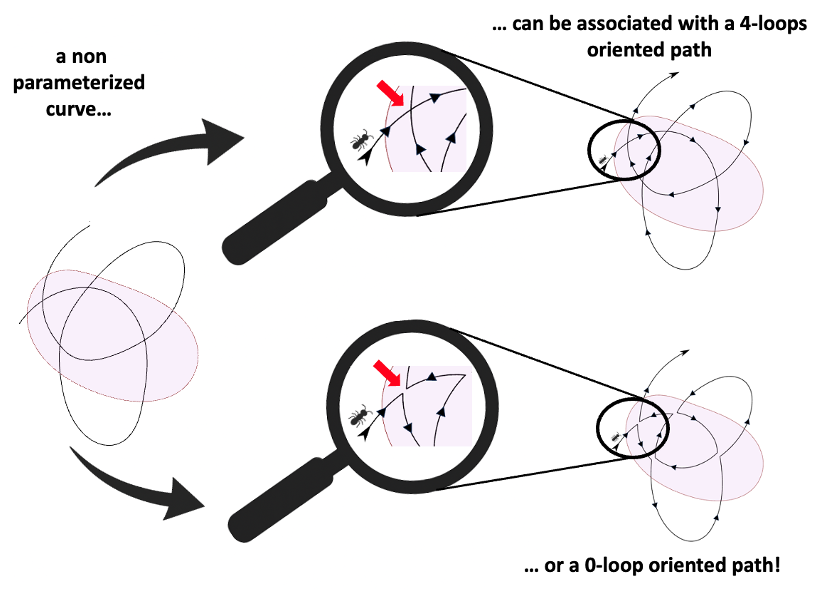}
  \setlength{\abovecaptionskip}{15pt}
  \caption{%
    Right: an experimental ant trajectory containing four loops, two of which lie inside the observation area.
    The path can be converted into a loop‑free curve (bottom‑right inset) by selecting, at each node, a branch that bypasses the crossing. 
    In both representations, the number of in‑domain segments is~3.
    While the loop‑free construction makes the invariance property manifest, it is not practically available, so the segment count must be obtained differently.  
    A graph‑theoretic analogy provides the required counting; note that the loop‑free decomposition is not unique.}
  \label{fig:ant}
\end{figure*}

\paragraph*{Integral-geometric strategy.}
The assumption of translational and rotational invariance (formally the Euclidean motion group in $n$ dimensions) is crucial: it allows us to invoke the kinematic formula of integral geometry, which underpins the generalized invariance property derived below.

We first treat a single random curve of fixed length~$L$\,\cite{footnote_2}.  We view the fixed observation window $K_0$ and the moving curve $K_1$ as two distinct geometric objects.  Evaluating the average in‑domain length is equivalent to integrating over all rigid motions that bring $K_1$ into contact with $K_0$.  We denote by $dK_1$ the kinematic density of $K_1$ (i.e., intuitively, $dK_1$ takes into account all the possible translations (uniform) and orientations (isotropic) of $K_1$).

Santal\'o showed that, for a fixed $n$‑dimensional manifold $K_0$ of volume $V_0$ and a moving one‑dimensional manifold $K_1$ of length $L$, the integral of the one‑dimensional volume of their intersection is~\cite{Santalo}
\begin{equation}
\label{L_nD_fixed}
  \int_{K_1\cap K_0\neq\varnothing} \ell\,dK_1 = O_{n-1}\cdots O_1\,V_0\,L ,
\end{equation}
where $O_m = 2\pi^{(m+1)/2}/\Gamma\!\bigl((m+1)/2\bigr)$ is the surface area of the unit $m$‑sphere.  
Equation~\eqref{L_nD_fixed} provides the first building‑block for the generalized IP.

The second building‑block—the number of paths within the observation region—is not directly provided by the integral geometry. However, Blaschke’s kinematic formula—originally established in two dimensions and later extended to higher dimensions by Santal\'o (see also Chern~\cite{Chern} and Chirikjian~\cite{Chirikjian})—provides
\begin{align}
\label{kinematic_formula_n_euclidean}
 & \displaystyle\int_{K_0 \cap K_1 \ne \varnothing}  \! \! \! \! \! \! \! \!  \! \! \! \!  \chi(K_0 \cap K_1) dK_1  = O_{1} \dots O_{n-2} {\Big[} O_{n-1} \Big( \chi(K_0) V_1  \nonumber \\
             & \qquad  \quad + \chi(K_1) V_0 \Big) + \frac{1}{n} \sum_{h=0}^{n-2} \binom{n}{h+1} M^0_{h} M^1_{n-2-h} {\Big]} 
\end{align}
where $\chi(K_0)$ and $\chi(K_1)$ denote the Euler characteristics of $K_0$ and $K_1$, respectively, $V_1$ the (zero) volume of $K_1$, and $\chi(K_0 \cap K_1)$ the Euler characteristic of their intersection.
The quantities $M_i^0$ and $M_i^1$ represent the $i$th integrals of the mean curvature of $K_0$ and $K_1$, respectively~\cite{deLin} (see Supplemental Material).
Equation~\eqref{kinematic_formula_n_euclidean} constitutes the second key ingredient in establishing the generalized Invariance Property. 
In the case of a simple (i.e., non-self-intersecting) trajectory, for which $\chi(K_1) = 1$, the Euler characteristic of the intersection $\chi(K_0 \cap K_1)$ precisely counts the number of distinct pieces of $K_1$ lying within $K_0$. If a trajectory is ``broken'' into multiple parts within $K_0$ 
(e.g., in Fig.~(\ref{fig-K0andK1-crop}) we have 2 pieces),
each piece is counted individually and contributes to the total number of pieces. 
The ratio of Eqs.\eqref{L_nD_fixed} and \eqref{kinematic_formula_n_euclidean} leads to the result (see Supplemental Material\ref{sec:suppProof_IP} for technical details)
\begin{equation}
\label{IP_fixed_length}
  \displaystyle  {\frac{1} {\langle \ell\rangle} = \frac{1}{L}+ \frac{1}{\langle \sigma \rangle} }\, .
\end{equation}
where $\langle\sigma\rangle = \eta_n V_0/S_0$ is the mean chord length with
\begin{equation}
\label{def_eta_n}
  \eta_n = \frac{(n-1)O_{n-1}}{O_{n-2}}
         = \sqrt{\pi}\,(n-1)\,\frac{\Gamma\!\bigl(\tfrac{n-1}{2}\bigr)}{\Gamma\!\bigl(\tfrac{n}{2}\bigr)}.
\end{equation}
Note that for infinite curves, $\langle \ell\rangle = \langle\sigma\rangle$ regardless of path shape.

So far, we have considered a single trajectory of length $L$. Since the result expressed in Eq.~\eqref{IP_fixed_length} depends only on $L$, it can be immediately extended to any type of trajectory, regardless of its shape, as long as they share the same length. 

\paragraph*{Beyond fixed-length, loops and branching.}
Besides,
a direct extension of the aforementioned result pertains to branching processes. Indeed, the sole hypothesis leading to Eq.\eqref{IP_fixed_length} is that the Euler characteristic of $K_1$ is equal to 1, which implies that the Euler characteristic of the intersection $\chi(K_0 \cap K_1)$ is the number of pieces of $K_1$ contained within $K_0$.
Now, a classical result of algebraic topology states that all contractible spaces have an Euler characteristic equal to 1~\cite{book_Zeidler}, which includes in particular curves with branches. Consequently, the Eq.\eqref{IP_fixed_length} remains valid for curves with ramifications (without loops) whose total length is $L$.

Finally, we consider the general case where the curve length $L$ is not fixed but drawn from an arbitrary probability distribution. Denoting by $\langle L \rangle$ the average curve length, when it exists, we obtain (see Supplemental Material for the detailed derivation):
\begin{equation}
\label{eq_cauchy_nD_random_final}
  \displaystyle  {\frac{1} {\langle \ell\rangle} = \frac{1}{\langle L \rangle}+ \frac{1}{\langle \sigma \rangle}} \, .
\end{equation} 
Equation~\eqref{eq_cauchy_nD_random_final}, which is the central result of this Letter, generalizes Cauchy’s formula to trajectories whose lengths follow an arbitrary probability distribution. 
It also generalizes earlier work by Kellerer\,\cite{Kellerer}, limited to straight segments in three dimensions.
A direct consequence of this result is that when the mean length of the trajectories diverges, Cauchy's formula holds:
\begin{equation}
\label{eq_cauchy_2D_infinite_curve_random}
  \displaystyle  \langle \ell\rangle = \langle \sigma \rangle = \eta_n \frac{V_0}{S_0} \mathrm{~~~when~} \langle L \rangle \rightarrow \infty \, ,
\end{equation}
regardless of the shape of the trajectories.

Note that a $K_1$ trajectory may be composed of several sub-trajectories. Even in this case Eqs.~(\ref{IP_fixed_length}) and (\ref{eq_cauchy_nD_random_final}) are valid. The only general condition is that each trajectory (sub-trajectory) has distinct beginning and ends. This means that {\it purely} circular trajectories (e.g. a circle) are not a suitable choice for $K_1$.

\emph{The 2-dimensional case} ---
First, observe that if a two-dimensional trajectory contains no loops, the reasoning previously outlined for $n \ge 3$ still applies, and the generalized IP remains valid. However, in two dimensions, random trajectories are prone to forming loops, which require careful handling.

For example, a single-loop trajectory has an Euler characteristic of zero. If this loop lies {\it entirely} within $K_0$, the Euler characteristic $\chi(K_0 \cap K_1)$ of the intersection between $K_0$ and $K_1$ also vanishes. As a result, the trajectory does not contribute to Eq.~\eqref{kinematic_formula_n_euclidean} and is effectively ignored.
More generally, trajectories with multiple loops are problematic. A trajectory with $k$ loops has Euler characteristic $1 - k$ and may contribute negatively to the integral in Eq.~\eqref{kinematic_formula_n_euclidean}.

\paragraph*{Loops re-interpreted as planar graphs.}
To address these issues, we first focus on simple, non-branching curves. As shown in Fig.~\ref{fig:ant},
any looped trajectory can be systematically reduced to an equivalent loop-free curve by selecting, at each node, a branch that avoids self-intersections.
This result follows directly from our assumption that the trajectory has a well-defined start and end point --- either on  $\partial K_0$ or in the interior of $K_0$  --- and from an analogy with graph theory. By interpreting a two-dimensional trajectory as a walk on a graph (i.e., a sequence of vertices and edges), a classical result applies: if a walk revisits the same vertex, which corresponds here to a self-intersection of the trajectory, it can be shortened by removing the internal loop, yielding a simpler, loop-free path~\cite{book_Diestel}.
This confirms again the validity of Eqs.~(\ref{IP_fixed_length}) and (\ref{eq_cauchy_nD_random_final}) in the case of non-branching trajectories with loops.

\paragraph*{Branching trajectories.}
In contrast, for branching trajectories with loops, this reduction no longer applies, and a different approach is needed to count the relevant paths in $K_0$.

\begin{figure}[h]
\centering
\includegraphics[width=2.5in,height=1.6in]{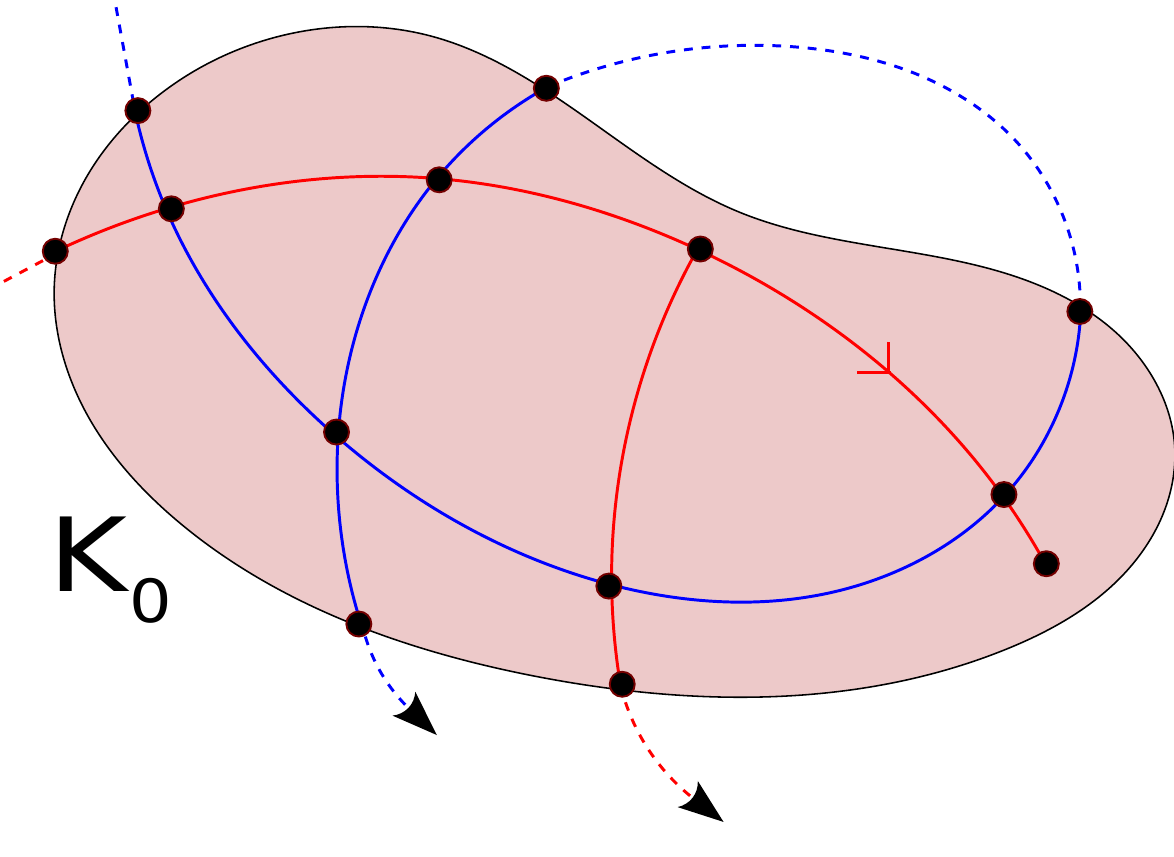}
\setlength{\abovecaptionskip}{15pt}  
\caption{Example of 2D traces where two trajectories — one of which (in red) branches and terminates within the domain — give rise to three independent paths ($N$) in the observation region.
In two dimensions, such traces can be interpreted as a graph. In this example, the graph consists of 13 vertices ($v$), indicated by black circles, including 5 4-branch vertices ($v_4$), and 15 edges ($e$), confirming the relation $N = v - e + v_4 = 3$, as expected.}
\label{fig:graph}
\end{figure}

If an observer can record passages through $K_0$ over time, counting is straightforward. But when only spatial traces are available, loops make the counting nontrivial.
As shown in the Supplemental Material, such a trajectory can be mapped onto a graph (see Fig.~\ref{fig:graph}).
A simple result then emerges: the number of independent trajectories $N$ within $K_0$ is given by $N = v - e + v_4$, where $v$ and $e$ are the number of vertices and edges in the graph, and $v_4$ is the number of 4-branch vertices (corresponding to crossings). This formula remains valid for branching trajectories, provided that no branching produces exactly three descendants, as detailed in the Supplemental Material.
This alternative counting method supports the validity of Eqs.~(\ref{IP_fixed_length}) and (\ref{eq_cauchy_nD_random_final}) even in the presence of loops and branching.

\emph{Numerical results and algorithm} ---
To test the universality of Eq.~\eqref{eq_cauchy_nD_random_final} we
performed large-scale Monte-Carlo (MC) simulations for four
different families of curves, summarized in~\cite{Santalo}
Fig.~\ref{fig2dand3d}:
\begin{figure*}[htbp]
    \centering
    \subfigure[needles uniformly distributed]{%
        \includegraphics[width=0.49\linewidth]{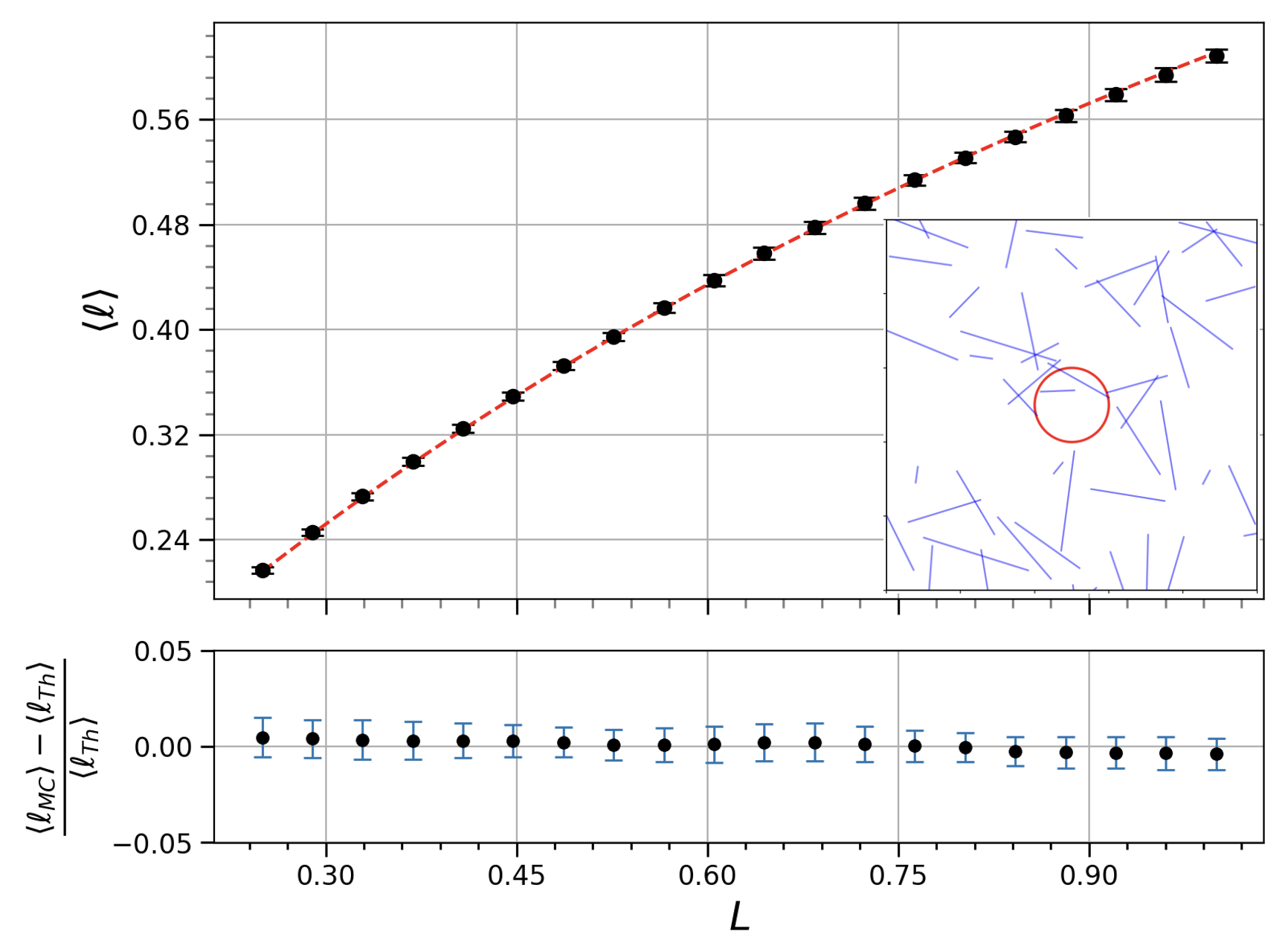}}
    \hfill
    \subfigure[Y-shaped]{%
        \includegraphics[width=0.5\linewidth]{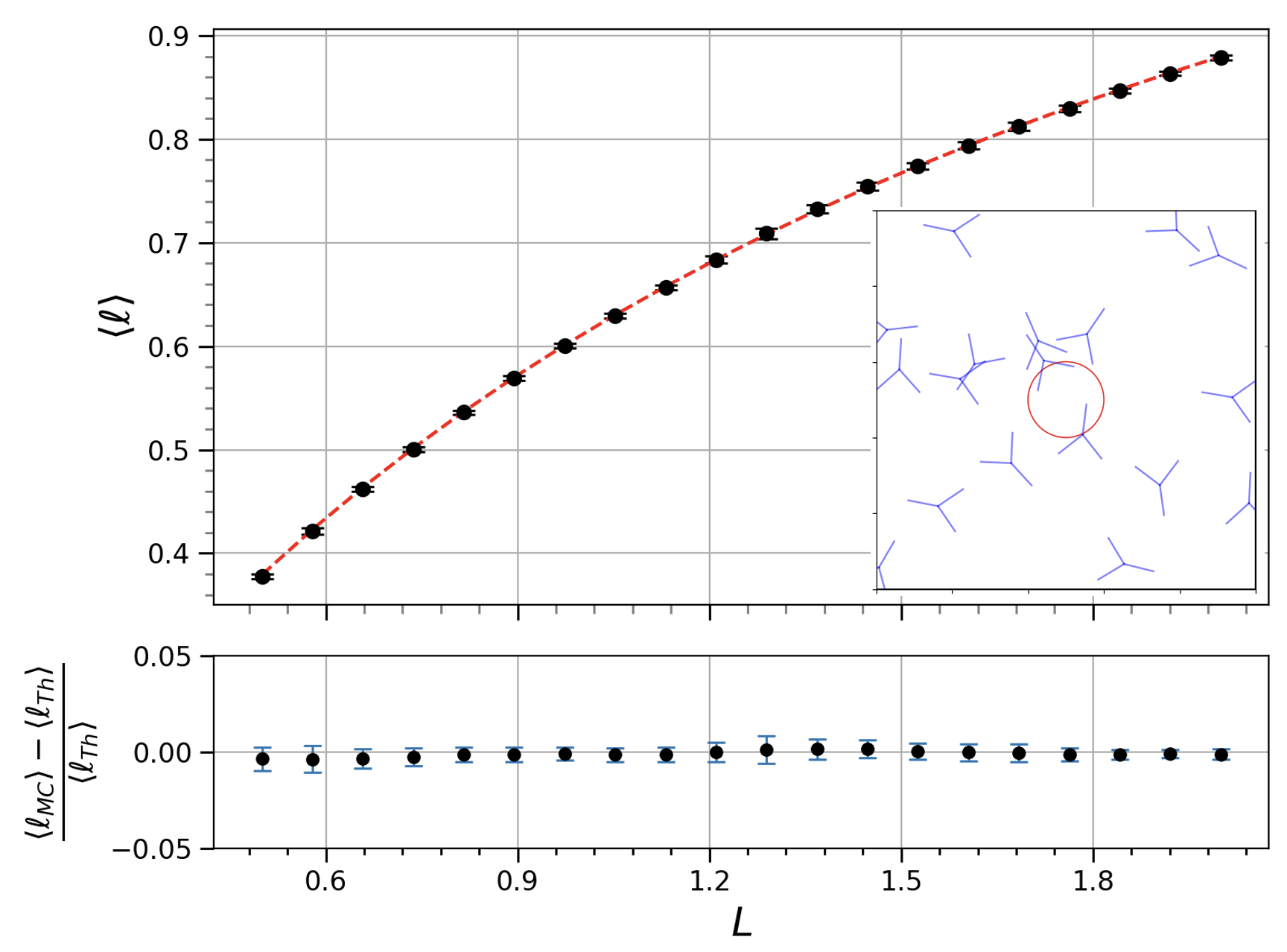}}

    \vspace{0.3cm}

    \subfigure[open triangle]{%
        \includegraphics[width=0.51\linewidth]{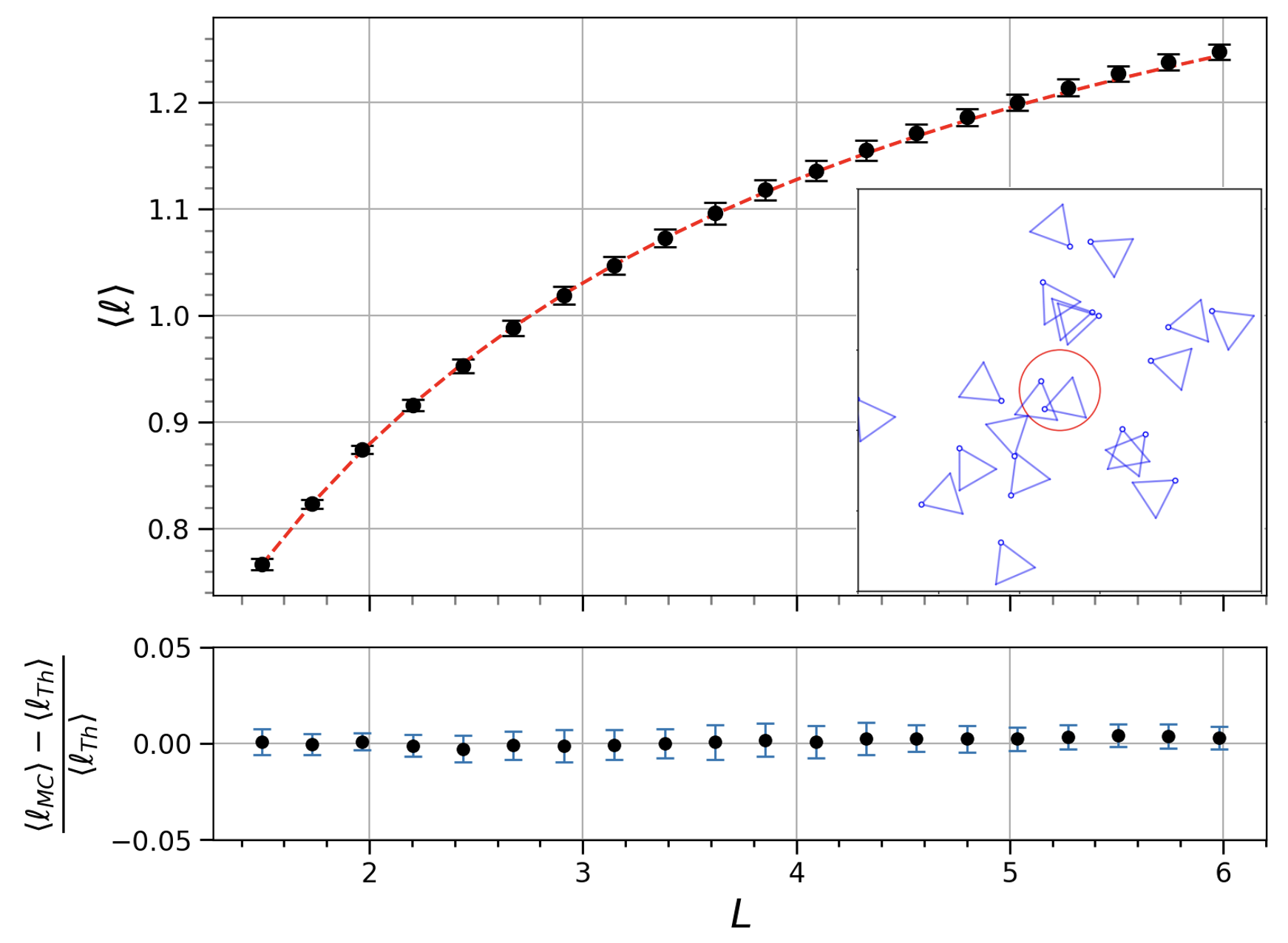}}
    \hfill
    \subfigure[isotropic random walk]{%
        \includegraphics[width=0.47\linewidth]{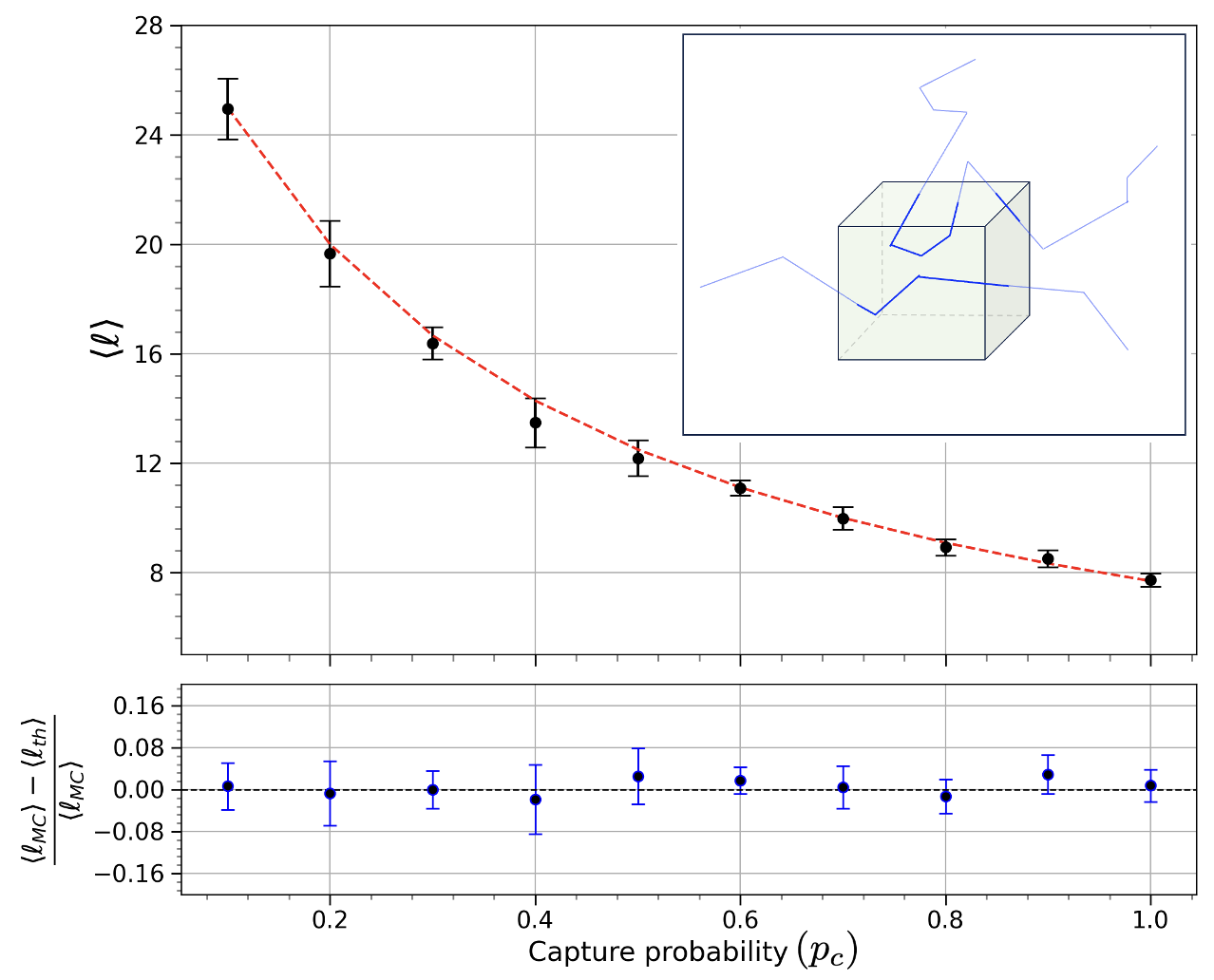}}

    \caption{Comparison between the generalized IP, Eq.~\ref{eq_cauchy_nD_random_final} (red dashed lines), and Monte Carlo simulations (black filled circles) for various configurations. Distances $\ell$ traveled within the detectors are in arbitrary units; relative differences are given in absolute values. Error bars represent three standard deviations of the mean.
(a) Distance $\langle \ell \rangle$ traveled within a disk-shaped detector by 2D straight needles of random length, uniformly distributed. Simulations used $10^6$ needles in a square domain (half-side 50 a.u.) with a disk detector of radius 1 a.u.
(b) Same comparison for 2D 'Y-shaped' random curves, using $10^6$ samples under identical geometric conditions.
(c) Same as (b), but for 2D \textit{open} triangles (marked by blue circles at one angle). Since loops are not permitted under the generalized IP, only open shapes are included.
Again, $10^6$ samples were used.
(d) Comparison of $\langle \ell \rangle$ versus capture probability $p_c$ for a 3D isotropic random walk with fixed jump size $a = 10$ a.u. Simulations used $10^7$ walks in a cubic domain (half-side 1000 a.u.) with a cubic detector of half-side 25 a.u.}
    \label{fig2dand3d}
\end{figure*}

\begin{enumerate}
\item[(a)] randomly oriented straight needles of length $L$ uniformly
      distributed in a prescribed interval (2D);
\item[(b)] fixed-length ``Y’’ shapes, i.e.\ a central stem of length
      $L/2$ with two identical branches, $60^{\circ}$ apart (2D);
\item[(c)] open triangles of perimeter $L$ (loops are forbidden by the
      theory) (2D);
\item[(d)] isotropic random walks in three dimensions with fixed step
      size $a$ and independent capture probability $p_c$ at each
      collision, so that the total walk length is stochastic (3D).
\end{enumerate}
\paragraph*{Simulation set-up.}
A very large “universe’’ volume is first defined—
a square of half-side~50 in 2D, a cube of half-side~$10^3$ in 3D—
inside which $N_{\text{curves}}$ trajectories are generated with
uniform random center and orientation.  
The much smaller detector $K_0$
(disk of radius~1 or cube of half-side~25, respectively) is placed
at the origin to ensure translational and rotational invariance
away from boundaries.

\begin{figure}
    \centering
    \includegraphics[width=2.5in,height=2.5in]{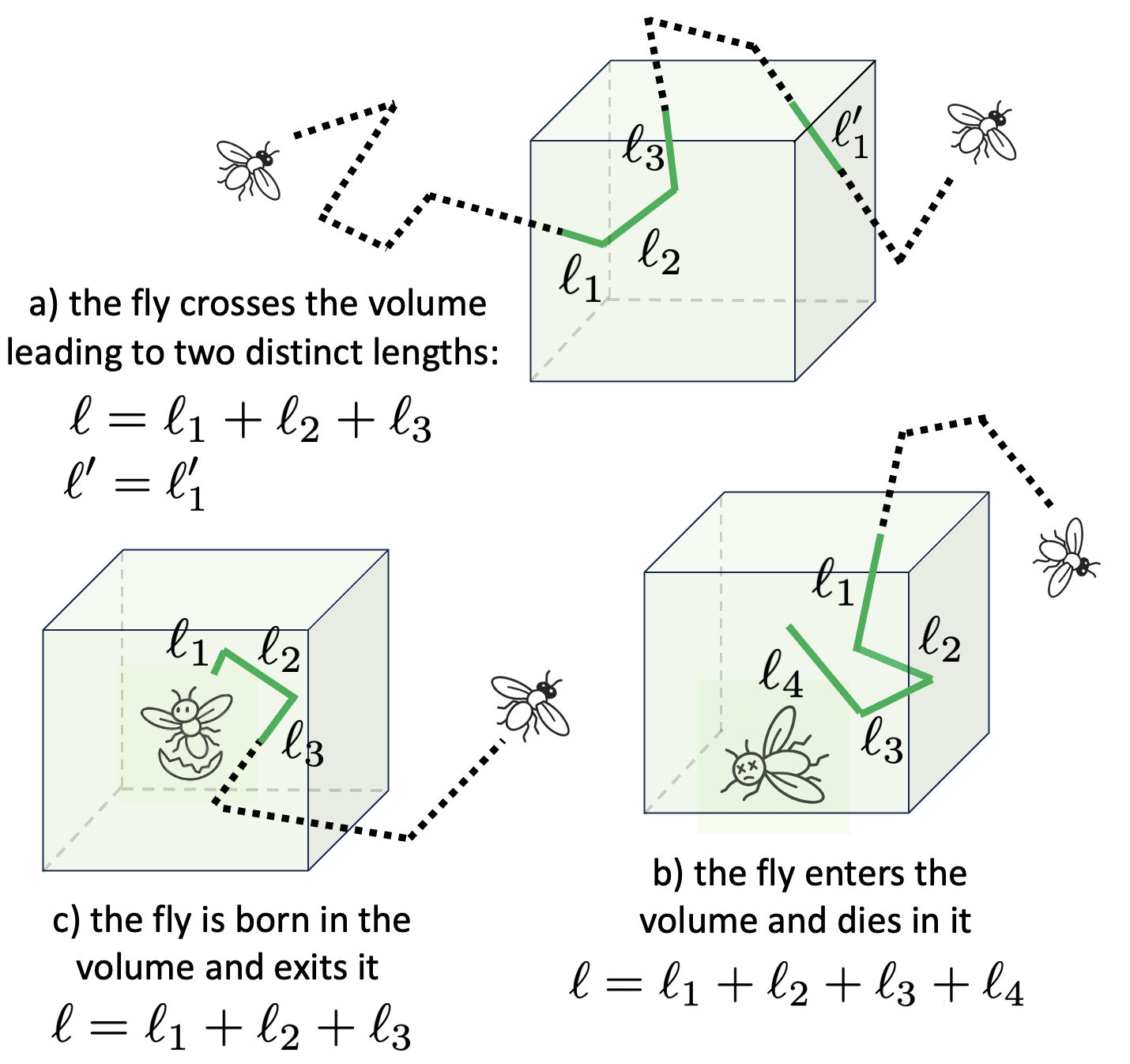}
    \setlength{\abovecaptionskip}{15pt} 
    \caption{The mean path length $\langle \ell \rangle$ of the fly through the cube —including trajectories that start, end, or pass through the volume— is given by the generalized IP formula in 3 dimensions: $1 / \langle \ell  \rangle = 1 / \langle L \rangle+ 1/{(4 V/S)}$, where $\langle L \rangle$ denotes the mean total length of the fly’s trajectory.}
    \label{fig:fly}
\end{figure}

Each curve is tracked segment-by-segment through $K_0$.
Whenever the trajectory enters the detector the partial
intersection length~$\ell$ is accumulated; if the curve exits and
re-enters, $\ell$ is reset and the new piece is tallied
independently, exactly as prescribed by the theory
(Fig.~\ref{fig:fly}). 
In the 3D random-walk case the algorithm also stores the realized
total length $L$ of every walk.

Statistics were gathered with
$10^{6}$–$10^{7}$ trajectories per data point; error bars in
Fig.~\ref{fig2dand3d} show three standard deviations of the mean.

\paragraph*{Results.}
Panels (a)–(c) in Fig. \ref{fig2dand3d} plot the measured $\langle \ell \rangle$ versus
$L$ for the three 2D curve families, together with the prediction
of Eq.~\eqref{eq_cauchy_nD_random_final} (dashed red lines).  
Panel (d) displays $\langle\ell\rangle$ versus the capture
probability $p_c$ for the 3D random walk, where
$\langle L\rangle=a/p_c$ is known analytically; once again the MC
points (black circles) sit precisely on the theoretical curve.


\emph{Conclusion} --- 
A purely geometric viewpoint reveals that the classic
\emph{invariance property} is only the infinite–length limit of a far broader law: for \emph{any} randomly placed and oriented curve, Eq.\eqref{IP_fixed_length} (or its averaged form,~Eq.\eqref{eq_cauchy_nD_random_final}) rigidly links the local mean length $\langle\ell\rangle$
measured in a detector to the global mean length
$\langle L\rangle$, provided their spatial distribution is uniform and isotropic.

In two dimensions, we complemented integral geometry with a 
minimal graph-theoretic counting argument, enabling a practical evaluation of 
the number of independent trajectory paths within the domain---even in the presence of loops and crossings---while excluding the singular case of ternary branchings.  

Because no dynamical assumptions enter the proof, the result applies
across scales: it is equally valid for photons in biological tissue or turbid media~\cite{book_Tiziano},
microorganismes such as prokaryotes~\cite{Pietrangeli}, 
and cosmic-ray cascades in the upper
atmosphere~\cite{Blanco_EPL,book_Goody}.  Wherever one can monitor \textit{only} a finite
window, Eqs.\eqref{IP_fixed_length}–\eqref{eq_cauchy_nD_random_final} turn that local snapshot into a global
diagnostic—an economical shortcut we expect to prove useful well
beyond the examples treated here.



\newpage
\begin{widetext}

\vspace{1.5cm}
\begin{center}
\hskip 0cm {\Large{Supplemental Material}}
\end{center}

\vspace{0.5cm}

\section{I. Detailed proof of the generalized Invariance Property}
\label{sec:suppProof_IP}
To derive the generalized Invariance Property we first need the
one‑dimensional mean‑curvature integrals $M_i^{\,1}$ for a curve.  We
start with a straight segment of length $s$ and then extend the result
to arbitrary curves.

For a line segment in $\mathbb{E}^n$ (i.e., $\mathbb{R}^n$ equipped with the Euclidean norm), Santal\'o~\cite{Santalo} gives

\begin{equation}
\label{mean_curvature_curve}
  \left\{
  \begin{aligned}
    M_i^{\,1} &= 0 , && i = 1,2,\dots , n-3, \\
    M_{n-2}^{\,1} &= \dfrac{O_{n-2}}{n-1}\,s, \\
    M_{n-1}^{\,1} &= O_{n-1}.
  \end{aligned}\right.
\end{equation}

Because the mean‑curvature integral is invariant under
bending~\cite{Almgren}, these values also hold for any rectifiable
curve of length $L$ in $\mathbb{E}^n$.  Substituting them into the
kinematic formula
[Eq.~\eqref{kinematic_formula_n_euclidean}] and noting that a curve
has zero volume ($V_1=0$) yields

\begin{equation}
  \int_{K_0 \cap K_1 \neq \varnothing}
  \chi\!\left(K_0 \cap K_1\right)\,dK_1
  = O_1\!\dots O_{n-2}
    \Bigl[
      O_{n-1}V_0
      + \frac{1}{n}\binom{n}{1} M_0^{\,0} M_{n-2}^{\,1}
    \Bigr].
\end{equation}

Since $M_0^{\,0}=S_0$, the surface area of $\partial K_0$, we obtain

\begin{equation}
\label{kinematic_formula_n_euclidean_curve}
  \int_{K_0 \cap K_1 \neq \varnothing}
  \chi\!\left(K_0 \cap K_1\right)\,dK_1
  = O_1\!\dots O_{n-2}
    \Bigl[
      O_{n-1}V_0
      + S_0\,\dfrac{O_{n-2}}{n-1}\,L
    \Bigr].
\end{equation}

Because $\chi(K_0 \cap K_1)$ counts each connected piece of $K_1$
inside $K_0$—including segments that start or end within
$K_0$ (Fig.~\ref{fig:fly})—the left‑hand side measures precisely that
number.  The average length $\langle \ell\rangle$ of $K_1$ inside $K_0$ is
therefore

\begin{equation}
  \langle \ell\rangle
  = \frac{%
       \displaystyle\int_{K_1 \cap K_0 \neq \varnothing} \ell\,dK_1}
       {%
       \displaystyle\int_{K_1 \cap K_0 \neq \varnothing}
       \chi\!\left(K_0 \cap K_1\right)\,dK_1}
  = \frac{%
       O_{n-1}\!\dots O_1\,V_0\,L}{%
       O_{n-2}\!\dots O_1
       \left[
         O_{n-1}V_0
         + \dfrac{O_{n-2}}{n-1}S_0L
       \right]}.
\end{equation}

Introducing the mean chord length 
$\langle\sigma\rangle
  =(n-1)(O_{n-1}/O_{n-2})(V_0/S_0)
  \equiv \eta_n V_0/S_0$
gives the compact form quoted in the main text
[Eq.~\eqref{IP_fixed_length}]:

\begin{equation}
\label{IP_fixed_length_bis}
  \frac{1}{\langle \ell\rangle}
  = \frac{1}{L} + \frac{1}{\langle\sigma\rangle}.
\end{equation}

\medskip
\noindent\textbf{Random‑length trajectories.}
Up to this point the trajectory length $L$ was fixed.  We now let $L$
be a random variable drawn from the density $f(L)$, and denote its expectation, if it exists, by $\langle L \rangle$. Compared to the fixed-length case, we must now average over all possible trajectory lengths according to the distribution \( f(L) \).
Averaging over both the kinematic density and the
length distribution, Eq.~\eqref{kinematic_formula_n_euclidean_curve}
becomes

\begin{align}
\label{eq_kinematic_random}
  \int_0^{\infty}\!dL\,f(L)
  \int_{K_1 \cap K_0 \neq \varnothing}
       \chi\!\left(K_1 \cap K_0\right)\,dK_1
  &= O_1\!\dots O_{n-2}
     \int_0^{\infty}\!dL\,f(L)
     \bigl[
       O_{n-1}V_0
       + S_0\,\frac{O_{n-2}}{n-1}L
     \bigr] \nonumber\\
  &= O_1\!\dots O_{n-2}
     \bigl[
       O_{n-1}V_0
       + S_0\,\frac{O_{n-2}}{n-1}\langle L\rangle
     \bigr].
\end{align}

Similarly,

\begin{equation}
\label{L_nD_random}
  \int_0^{\infty}\!dL\,f(L)
  \int_{K_1 \cap K_0 \neq \varnothing} \ell\,dK_1
  = O_{n-1}\!\dots O_1\,V_0\,\langle L\rangle.
\end{equation}

Taking the ratio again gives the average length $\langle \ell \rangle$ of $K_1$ inside $K_0$

\begin{equation}
\label{eq_cauchy_nD_random_final_bis}
  \frac{1}{\langle \ell \rangle}
  = \frac{1}{\langle L \rangle}
    + \frac{1}{\langle \sigma \rangle},
\end{equation}

in agreement with Eq.~\eqref{eq_cauchy_nD_random_final} of the main text.

\section{II. Two dimensional case: number of paths in the observation zone}
\label{sec:supp2D}
In this paragraph, we establish—via an analogy with graph theory—that in the two-dimensional case, and regardless of whether the trajectories contain loops, the number of distinct paths \( N \) within the observation domain is given by  
\begin{equation}
\label{eq_N}
    N = v - e + v_4,
\end{equation}  
where \( v \), \( e \), and \( v_4 \) denote the number of vertices, edges, and 4-branch vertices, respectively, in the graph formed by the intersections of the trajectories with \( K_0 \), as illustrated in Fig.~\ref{fig:graph}.

We begin by observing that the number of paths is equal to the sum of the Euler characteristic \( \chi(K_0 \cap K_1) \) and the number of 4-branch vertices within \( K_0 \). This relation can be demonstrated by induction.

We first consider a simple, loop-free trajectory intersecting \( K_0 \), for which \( \chi(K_0 \cap K_1) = 1 \). In this case, there are no 4-branch vertices, and clearly \( N = \chi(K_0 \cap K_1) = 1 \). Now, suppose a second trajectory intersects \( K_0 \), so that \( N = 2 \). If this new path does not intersect the first, then \( \chi(K_0 \cap K_1) = 2 \), and again \( N = \chi(K_0 \cap K_1) \), since no 4-branch vertices are introduced. However, if the two trajectories intersect at a single point, the Euler characteristic becomes \( \chi(K_0 \cap K_1) = 1 \), and a single 4-branch vertex appears, yielding \( N = \chi(K_0 \cap K_1) + v_4 = 2 \). If the two paths intersect at two points, a loop is formed, reducing the Euler characteristic to \( \chi(K_0 \cap K_1) = 0 \), and two 4-branch vertices appear, again leading to \( N = \chi(K_0 \cap K_1) + v_4 = 2 \).

This reasoning holds regardless of the number of loops, as each new loop that decreases \( \chi(K_0 \cap K_1) \) by one is compensated by the appearance of a new 4-branch vertex. Thus,
\begin{equation}
N = \chi(K_0 \cap K_1) + v_4.
\end{equation}

In practice, directly computing the Euler characteristic \( \chi(K_0 \cap K_1) \) requires identifying loops—a nontrivial task. This relation can therefore be refined using a well-known identity from graph theory:  
\begin{equation}
\label{eq_relation_graph}
    \chi(K_0 \cap K_1) = v - e,
\end{equation}
(see, e.g., Ref.~\cite{book_Diestel}). Substituting into the previous expression yields the final result:  
\[
N = v - e + v_4.
\]

We now consider the case of branching paths, beginning with a scenario in which a particle gives rise to two offspring. This situation introduces 3-branch vertices, which do not affect the definition of \( N \) and thus leave the previous result unchanged.
  
In contrast, branching walks that produce \emph{three} offspring introduce a complication. As illustrated in Fig.~\ref{fig:4-vertex},
\begin{figure}[h]
\centering
\includegraphics[width=4.2in,height=1.7in]{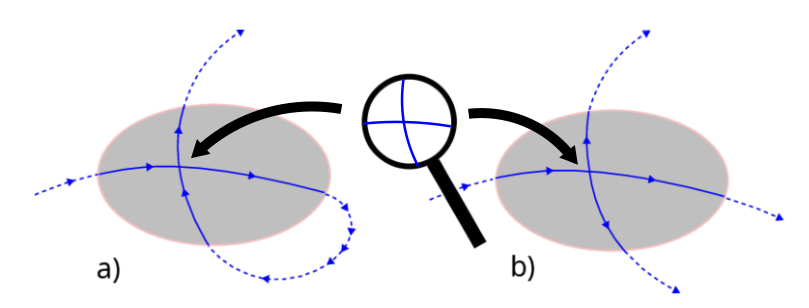}
\setlength{\abovecaptionskip}{15pt}  
\caption{Examples of 4-vertices: a) a 4-vertex resulting from the intersection of two paths, corresponding to two trajectory segments in \( K_0 \); b) a 4-vertex arising from a branching event with three descendants, corresponding to a single trajectory segment in \( K_0 \). Based solely on the observed traces within the domain, the two types of 4-vertices cannot be distinguished (see magnified view), which motivates the exclusion of branching events with three descendants.}
\label{fig:4-vertex}
\end{figure}
it becomes topologically and practically impossible to distinguish between a 4-branch vertex arising from the intersection of two trajectories—which corresponds to two path segments—and a 4-branch vertex resulting from a single trajectory branching into three descendants. This ambiguity is specific to the case of three offspring, as only in this case does a branching vertex contribute to the term \( v_4 \) in Eq.~\eqref{eq_N}. Note that such an ambiguity can only occur when two of the three outgoing trajectories point in precisely opposite directions—a very rare event—as illustrated by the upward and downward paths in Fig.~\ref{fig:4-vertex}b.

When a branching event produces more than three offspring, no ambiguity arises. For example, if the number of descendants is even, the corresponding vertex will have an odd number of branches—an event that cannot result from simple trajectory crossings. In addition, the chance that three or more independent trajectories intersect at exactly the same point is effectively zero, so such cases can be safely ignored.

We can therefore conclude that any vertex with more than four branches must result from a genuine branching event and does not affect the value of \( N \) given in Eq.~\eqref{eq_N}.

\end{widetext}


\begin{thebibliography}{99}
\bibitem{footnote_1}
Under this assumption, the number of trajectories entering the domain is proportional to the cosine of the angle between the inward normal and the trajectory direction (see Ref.~\cite{Bardsley}).
\bibitem{footnote_2} 
We assume that the one-dimensional object is rectifiable, even if it extends to infinite length. This criterion automatically excludes Brownian paths due to their fractal nature.
\end{thebibliography}
\end{document}